\documentclass[11pt]{article}
\usepackage{amsmath,amssymb,color,graphics,epsfig,cite}

\usepackage{amsfonts}

\newcommand{\hoch}[1]{$\, ^{#1}$}

\newcommand{\be}{\begin{equation}}
\newcommand{\ee}{\end{equation}}
\newcommand{\bea}{\setlength\arraycolsep{2pt} \begin{eqnarray}}
\newcommand{\eea}{\end{eqnarray}}
\newcommand{\nn}{\nonumber}

\def\ft#1#2{{\textstyle{\frac{\scriptstyle #1}{\scriptstyle #2} } }}
\def\fft#1#2{{\frac{#1}{#2}}}

\def\0{{\sst{(0)}}}
\def\1{{\sst{(1)}}}
\def\2{{\sst{(2)}}}
\def\3{{\sst{(3)}}}
\def\4{{\sst{(4)}}}
\def\5{{\sst{(5)}}}
\def\6{{\sst{(6)}}}
\def\7{{\sst{(7)}}}
\def\8{{\sst{(8)}}}
\def\sst#1{{\scriptscriptstyle #1}}

\begin{document}

\begin{center}
{\Large {\bf Black Hole Scalarization in Gauss-Bonnet Extended Starobinsky Gravity}
}

\vspace{20pt}

{\large Hai-Shan Liu\hoch{1}, H. L\"u\hoch{1}, Zi-Yu Tang\hoch{2} and Bin Wang\hoch{3,2}}

\vspace{10pt}

\hoch{1}{\it  Center for Joint Quantum Studies and Department of Physics,\\
School of Science, Tianjin University, Tianjin 300350, China}

\vspace{10pt}
\hoch{2}{\it School of Aeronautics and Astronautics, \\ Shanghai Jiao Tong University, Shanghai 200240, China}

\vspace{10pt}
\hoch{3}{\it Center for Gravitation and Cosmology, Yangzhou University, Yangzhou 225009, China}

\vspace{40pt}

\underline{ABSTRACT}
\end{center}

We propose a class of higher-derivative gravities that can be viewed as the Gauss-Bonnet extension of the Starobinsky model.  The theory admits the Minkowski spacetime vacuum whose linear spectrum consists of the graviton and a massive scalar mode. In addition to the usual Schwarzschild black hole,  we use numerical analysis to establish that in some suitable mass range, new black holes carrying the massive scalar hair can emerge.  The new black hole serves as a ``wall'' separating the naked spacetime singularity and wormholes in the parameter space of the scalar hair. Our numerical results also indicate that although the new hairy black hole and the Schwarzschild have different spacetime geometry, their entropy and temperature are same for the same mass.

\vfill{\footnotesize  hsliu.zju@gmail.com \ \ \ mrhonglu@gmail.com \ \ \ tangziyu@sjtu.edu.cn \ \ \ wang\_b@sjtu.edu.cn }



\thispagestyle{empty}
\pagebreak

\section{Introduction}

A natural extension to Einstein's theory of General Relativity is to include higher order Riemann tensor polynomial invariants.  The simplest example is the quadratic extension and the theory was proven renormalizable in four spacetime dimensions, at the price of having massive spin-2 ghostlike excitations in the spectrum \cite{Stelle:1976gc}. The origin of ghost is due to the fact that the field equations can involve up to fourth-order derivatives, leading to new massive scalar and massive spin-2 modes in the linearized spectrum, in addition to the usual spin-2 massless graviton.  There have been various proposals to deal with the ghost excitations.  One is to consider critical values of the couplings such that the massive spin-2 mode becomes massless \cite{Li:2008dq,Bergshoeff:2009hq,Liu:2009bk,Lu:2011zk}, but it typically leads to logarithmic ghost modes \cite{Bergshoeff:2011ri,Porrati:2011ku}.  An alternative is to consider combinations of the Riemann tensor polynomials such that the field equations remain in second order.  These include the the Gauss-Bonnet or more general the Lovelock series, which are nontrivial only in dimensions $D\ge 5$. The absence of massive modes in the linearized spectrum in maximally-symmetric vacua can be achieved in four dimensions.  This class of massless gravity theories include quasi-topological gravities and their variants, e.g.~\cite{Oliva:2010eb,Myers:2010ru,Bueno:2016xff,Li:2017ncu}.

In higher derivative gravities, while the massive spin-2 modes are inevitably ghostlike, the massive scalar mode can be unitary.  The ghost free condition can thus be achieved by decoupling the massive spin-2 mode, which amounts to a single condition on the coupling constants in the theory. The simplest such an example is the celebrated Starobinsky $R + R^2$ model.  The theory admits the Schwarzschild black hole, but no static black holes can carry the scalar hair \cite{Nelson:2010ig,Lu:2015cqa}.  By contrast, new black holes with massive spin-2 hair do exist in quadratically-extended gravity \cite{Lu:2015cqa}. This leads to an obvious question whether the no-scalar-hair theorem is a general phenomenon in higher derivative gravities or only a special case for Starobinsky gravity.

The absence of the scalar hairy black holes in the Starobinsky model is analogous to the no-hair theorem in Einstein gravity minimally coupled to a free scalar.  Recently, it was shown by numerical approach that scalar hairy black holes can be constructed when the free  massless scalar field is further coupled to the Gauss-Bonnet term \cite{95esgb,esgb1,esgb2,esgb3,kesgb}.  Analogous construction for a massive scalar field was subsequently obtained \cite{Doneva:2019vuh}. These motivate us to consider extending the Starobinsky theory with the Gauss-Bonnet combination in an appropriate way such that its massive scalar mode can be excited by the black hole curvature.  The resulting theory remains pure gravity and ghost free, constructed from the Ricci scalar and the Gauss-Bonnet term.

The advantage to study the black hole scalarization in the Gauss-Bonnet extended Starobinsky model is that the theory is pure gravity and we do not need to introduce matter scalar that may not exist. In fact, the prevalence of the scalar mode in higher-derivative gravities makes the study of black hole scalarization an unavoidable topic. Furthermore, the scalar mode in the Starobinsky model is necessarily massive and its effect is invisible in the long range. The scalarization however implies that it can nontrivially affect the black hole horizon.

\section{A no-hair theorem and its evasion}

The simplest higher order extension to Einstein gravity is to include the quadratic curvature invariants. In four dimensions, owing to the fact that the Gauss-Bonnet combination is a total derivative, the quadratic invariants of the Riemann tensor have two independent structures, the squared Ricci scalar and Weyl tensor, which excite respectively a massive scalar mode and a massive spin-2 mode in the linear spectrum.  The extended gravity admits the usual Schwarzschild black hole; in addition, new back holes carrying massive spin-2 hair was constructed \cite{Lu:2015cqa}.  However, a no-scalar-hair theorem can be established, for which it is sufficient to consider only the $R^2$ extension, namely the Starobinsky model
\be
{\cal L}=\sqrt{-g} (R + \alpha R^2)\,,\qquad \alpha\ge 0\,.\label{stmodel}
\ee
The trace of the Einstein equation is simply $6\alpha \Box R=R$. Since $R$ acts like a free massive scalar field here, it is straightforward to prove that $R$ must vanish for black hole solutions \cite{Lu:2015cqa}.

The Starobinsky gravity is the simplest example of $f(R)$ gravity and it is an effective scalar-tensor theory where the scalar equation is algebraic.  We can equivalently express (\ref{stmodel}) as
\be
{\cal L}=\sqrt{-g} (R + \phi R - \ft12\mu^2 \phi^2)\,,\qquad \mu^2 = \fft{1}{2\alpha}>0\,.
\ee
Inspired by \cite{esgb1,esgb2,esgb3}, we extend the theory with the Gauss-Bonnet invariant, {\it i.e.}
\be
{\cal L}=\sqrt{-g} (R + \phi R - \ft12\mu^2 \phi^2 + U(\phi) E^{\rm GB})\,,\qquad E^{\rm GB}=R^2 - 4R^{\mu\nu}R_{\mu\nu} + R^{\mu\nu\rho\sigma} R_{\mu\nu\rho\sigma}\,.\label{genlag}
\ee
The Einstein equation is
\bea
&& R_{\mu\nu} - \ft12 R  g_{\mu\nu} + \phi R_{\mu\nu} - \nabla_\mu\nabla_\nu \phi + \Box \phi g_{\mu\nu} - \ft 12 \phi R g_{\mu\nu} + \ft 14\mu^2\phi g_{\mu\nu}-2 R\nabla_\mu\nabla_\nu U\nn\\
&&- 4 (R_{\mu\nu}-\ft12 R g_{\mu\nu}) \Box U + 8 R^\rho{}_{(\mu} \nabla_{\nu)}\nabla_\rho U - 4 R^{\rho\sigma}\nabla_\rho\nabla_\sigma U g_{\mu\nu} + 4  R_\mu{}^\rho{}_\nu{}^\sigma \nabla_\rho \nabla_\sigma U = 0 \,. \label{einsteineom}
\eea
Combining the algebraic scalar field equation with the trace of (\ref{einsteineom}), we have
\be
3\Box \phi =\mu^2 \phi - (1 + \phi) U'(\phi) E^{\rm GB} - 2 R \Box U(\phi) + 4 R^{\mu\nu} \nabla_\mu\nabla_\nu U.
\ee
When the coupling function $U$ vanishes, the equation describes a standard free massive scalar and the no-hair theorem applies for black hole solutions. The no-hair theorem is no longer applicable when the scalar-Gauss-Bonnet term is included, making it possible for scalar hairy black holes.

It is important that the scalar equation in the extended theory is algebraic and hence can be integrated out to give pure gravity.  We present two concrete examples.  For $U=\beta \phi$, we have $\phi=2\alpha (R +\beta E^{\rm GB})$ and hence
\be
{\cal L}=\sqrt{-g} \Big( R + \alpha (R + \beta E^{\rm GB})^2\Big)\,.\label{extension1}
\ee
For $U=\ft12 \beta \phi^2$, we have $\phi=2\alpha R/(1 - 2\alpha\beta E^{\rm GB})$ and hence
\be
{\cal L}=\sqrt{-g}\Big(R + \fft{\alpha R^2}{1-2\alpha \beta E^{\rm GB}}\Big)\,.\label{extension2}
\ee
Both Gauss-Bonnet extensions of the Starobinsky model are pure gravity theories.

\section{Black hole constructions}

We construct static and spherically-symmetric black hole solutions that are asymptotic to the Minkowski spacetime.  We consider the Lagrangian (\ref{genlag}) with $U=\ft12\beta \phi^2$, corresponding to the Starobinsky model extended with the Gauss-Bonnet invariant (\ref{extension2}). The general ansatz is
\be
ds^2 =- h(r) dt^2 + \fft{dr^2}{f(r)} + r^2 \big(d\theta^2 + \sin^2\theta\, d\varphi^2\big)\,,\qquad \phi=\phi(r)\,.
\ee
Since $\phi=0$ is clearly a solution, the theory admits the usual Schwarzschild black hole $h=f=1-2M/r$. Here we investigate whether there can exist new black holes that carry the massive scalar hair. At asymptotic infinity, the scalar has a Yukawa falloff
\be
\phi = \fft{\phi_0 }{r} e^{-\fft{\mu}{\sqrt3}r} + \cdots\,.
\ee
This modifies the metric functions, with leading falloffs
\be
h=1 - \fft{2M}{r} - \fft{\phi_0}{r} e^{-\fft{\mu}{\sqrt3} r} + \cdots\,,\quad
f=1 - \fft{2M}{r} + \phi_0\Big(\fft{1}{r}+ \fft{\mu}{\sqrt 3}\Big)e^{-\fft{\mu}{\sqrt3} r} + \cdots\,.
\ee
The ADM mass is $M$, independent of the scalar hair $\phi_0$, or the couplings $(\mu,\beta)$ of the theory. Consequently the massive scalar is effectively invisible in the long range. However, it can alter the horizon structure significantly. The leading near-horizon expansions are
\be
h=h_1 (r-r_+)+\cdots \,, \qquad  f= f_1 (r - r_+) + \cdots \,, \qquad \phi= \phi_+ + \phi_1 (r - r_+ )+ \cdots \,,
\ee
with
\bea
f_1 &=& \frac{\left(3 + 3\phi_+-\beta  \mu ^2 \phi_+^3\right)r_+}{12 \beta  \phi_+ (\phi_++1)}
 -\frac{1}{48 \beta  \phi_+\left(\phi_++1\right)}  \Big[ 16 r_+^2 \left(-\beta  \mu ^2 \phi_+^3+3 \phi_++3\right)^2  \cr
  && \quad -96 \beta  \phi_+ (\phi_++1) \left(4 \beta  \mu ^2 \phi_+^3-3 \mu ^2 \phi_+^2 r_+^2-2 \phi_+ \left(\mu ^2 r_+^2-6\right)+12\right) \Big]^{\fft12} \,,\nn\\
\phi_1 &=& \frac{4 (1-f_1 r_+) (1+\phi_+) - \mu ^2 \phi_+^2 r_+^2}{2f_1( 4\beta \phi_++r_+^2)} \,.
\eea
The horizon is characterized by two integration constants, the horizon radius $r_+$ and the horizon scalar hair $\phi_+$, matching the asymptotic $M$ and $\phi_0$. The constant $h_1$ can be in principle arbitrary owing to the time scaling invariance; it is determined by requiring the resulting $h$ approach unit asymptotically. We can then read off the temperature and entropy
\be
T=\fft{\sqrt{h_1 f_1}}{4\pi}\,,\qquad S = \pi r_+^2 (1 + \phi_+ ) + 2 \pi \beta \phi_+^2.
\ee

In the absence of exact solutions, we use a numerical approach to connect the horizon data $(r_+, \phi_+)$ to asymptotic ones $(M, \phi_0)$.  The theory itself contains two coupling parameters $(\mu, \beta)$ and they will be fixed to some appropriate values. One way is to integrate the solution from the horizon to large $r$. Since the equations are singular on the horizon, we can perform the Taylor expansions and shift the initial integration point slightly out from the horizon. However, the complexity of the expressions for $(f_1,\phi_1)$ implies that we cannot analytically push the Taylor series to much higher orders to improve the accuracy. Furthermore, generic $(r_+,\phi_+)$ values will in general excite the divergent $e^{+\mu r/\sqrt3}$ mode in the scalar, making the numerical analysis very unstable. Only the extremely fine-tuned balance between $(r_+,\phi_+)$ may lead to a black hole solution.  This is typical for solutions involving massive mode, as in the case of the hairy black holes constructed in \cite{Lu:2015cqa}.

\begin{figure}[htp]
\begin{center}
\includegraphics[width=320pt]{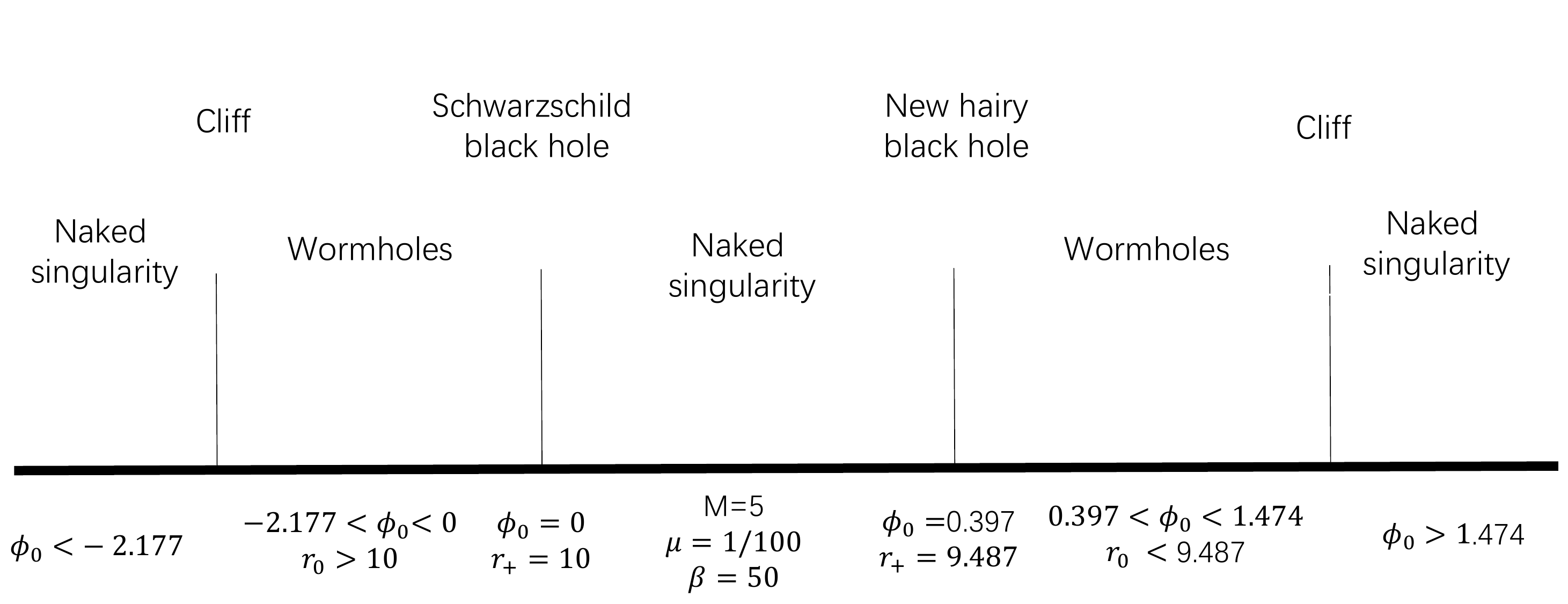}
\end{center}
\caption{\small\it This sketches the existence of a new scalar hairy black hole of mass $M=5$ for the $\mu=1/100$ and $\beta=50$ theory. The Schwarzschild black hole ($r_+^s=10$) is hairless ($\phi_0=0$) and a new black hole with $r_+=9.487$ emerges at $\phi_0=0.397$.  Both black holes are at the boundaries transiting from naked singularity to wormholes. Analogous transitions at $\phi_0=-2.177, 1.474$, however, are like cliffs and yield no black hole.}
\label{schematicfig}
\end{figure}

Alternatively, we can integrate the solution from large $r$ to the middle.  Since the scalar mode falls off exponentially, the higher-order corrections can be ignored numerically.  For appropriately chosen $(M,\phi_0)$ parameters, a black hole is characterized by the fact that the functions $(h,f)$ will vanish simultaneously somewhere that can be identified as the horizon. In practice, this is difficult to achieve precisely since the equations are singular on the horizon. One can nevertheless establish the existence of the scalar hairy black holes. For a concrete example, we consider $\mu=1/100$, $\beta=50$ and $M=5$. Its Schwarzschild radius is $r_+^s=10$. For negative values of $\phi_0$, we find that the function $f$ approaches zero at some $r_0\ge r_+^s$ before $h$, giving rise to a solution that describes half of a wormhole. The wormhole throat $r_0$ increases as $\phi_0$ becomes more negative. For positive and small $\phi_0$, we find that naked singularity develops where the function $h$ reaches zero at $r_0<r_+^s$, and $f$ diverges. As $\phi_0$ increases, $r_0$ decreases and the divergence of $f$ becomes less and less severe until $\phi_0$ reaches a critical value $\phi_0^{*}=0.397$ where $f$ and $h$ vanish simultaneously at $r_+=9.487$. This leads to a new scalar hairy black hole.  If we keep on increasing $\phi_0$, we find that $f$ approaches zero at smaller $r_0$ before $h$, giving rise to wormholes again.  Thus for appropriately given mass $M$, in the line of scalar hair $\phi_0$, the Schwarzschild and new black holes are ``walls'' separating wormholes from naked singularities. However analogous transitions at $\phi_0=-2.177$ and $\phi_0=1.474$ are like ``cliffs'' with no black holes. We sketch this in Fig.~\ref{schematicfig}.

\begin{figure}[htp]
\begin{center}
\includegraphics[width=190pt]{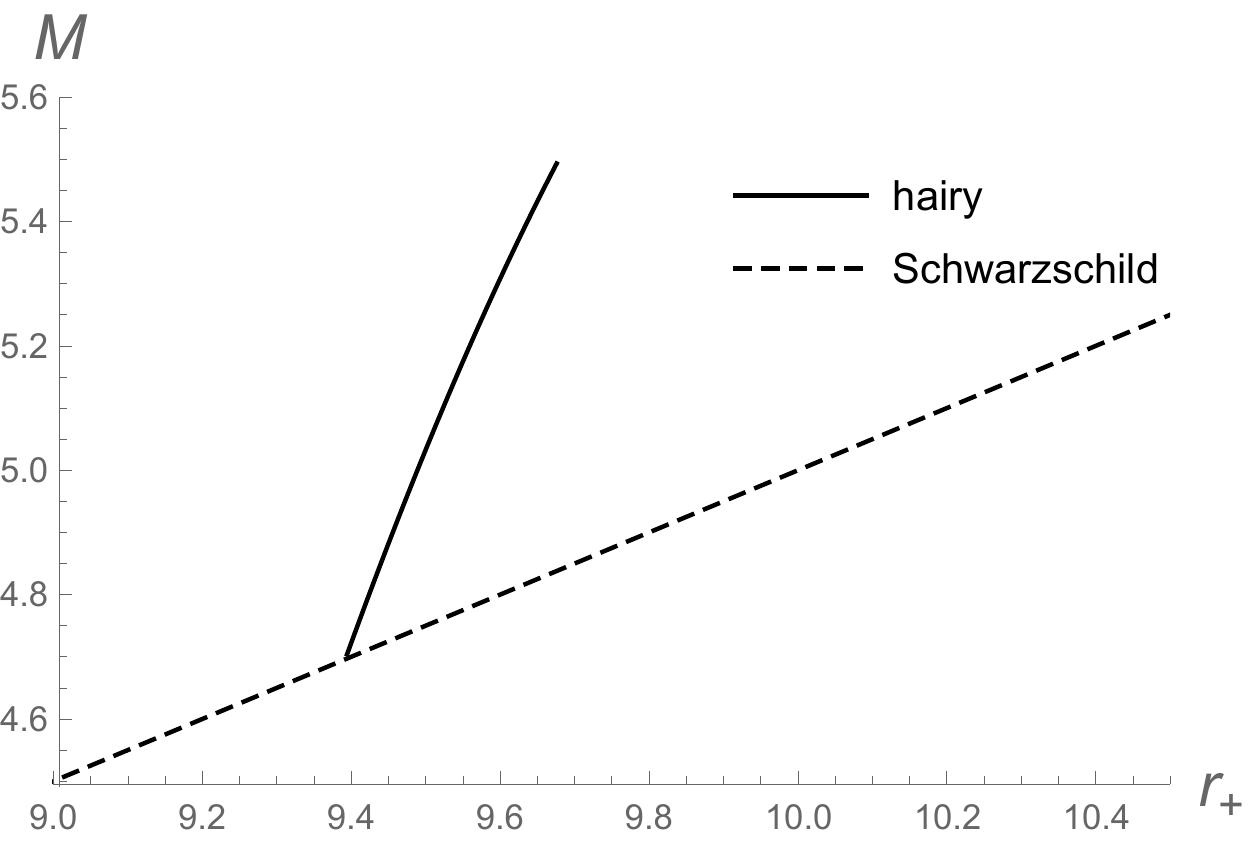}\ \ \ \ \
\includegraphics[width=210pt]{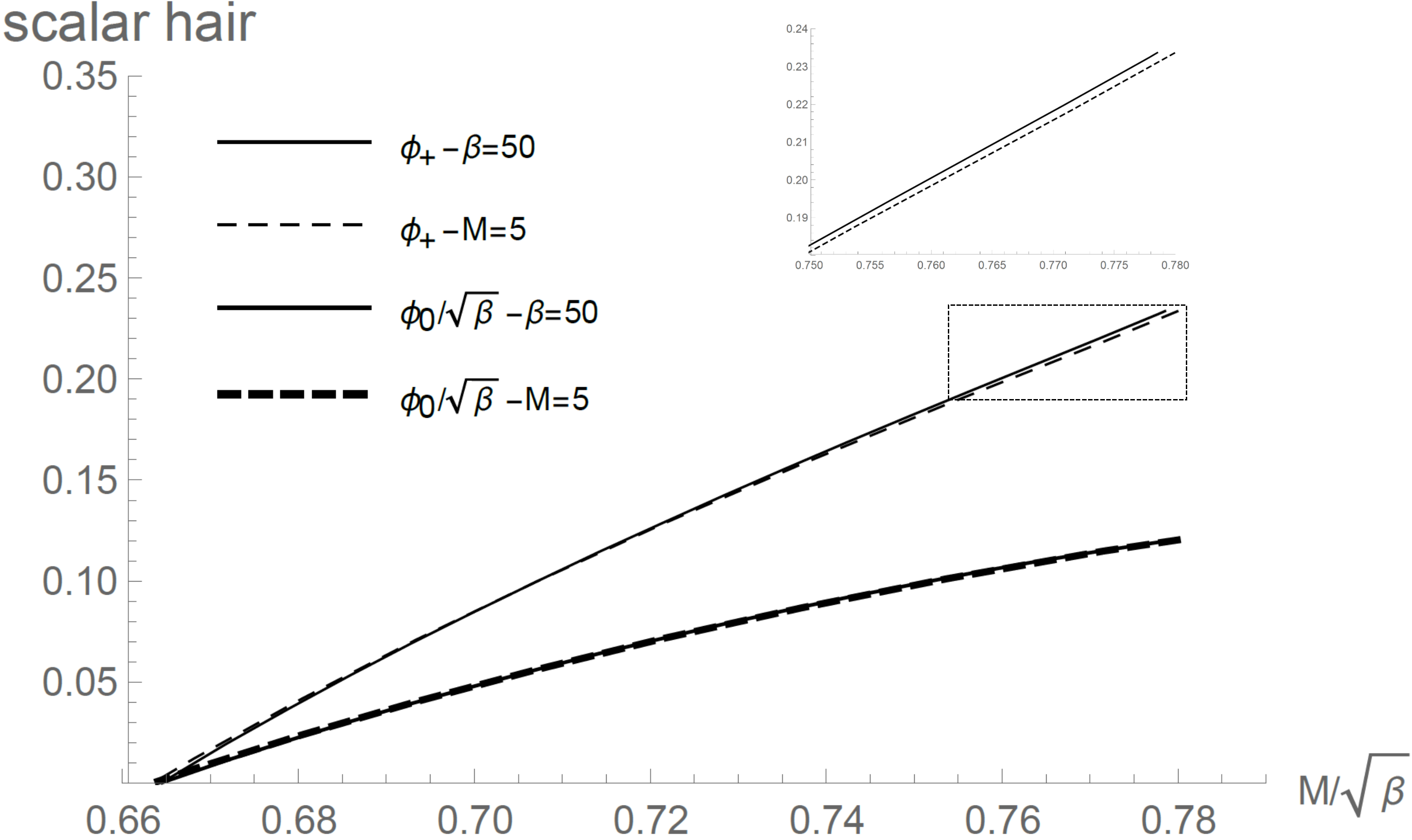}
\end{center}
\caption{\small\it  The left panel shows the range of the allowed mass $M$ for the new black holes with $(\mu=1/100,\beta=50)$ and their mass dependence on the horizon radius, compared to the Schwarzschild. The right panel shows that the (dimensionless) scalar hair parameters $(\phi_+,\phi_0/\sqrt{\beta})$ are not independent, but functions of $\tilde M=M/\sqrt{\beta}$, as well as either $\tilde \mu = \mu \sqrt{\beta}$ or $\hat \mu = \mu M$ for a different choice of parametrization. The dependence on $\mu$ can be seen by the slight difference between the lines of scalar hair for either fixed $\tilde \mu$ or $\hat \mu$. The solutions all have $\mu=1/100$, and hence the solid lines are for fixed $\tilde \mu$ and the dashed lines are for fixed $\hat \mu$.} \label{mrscalarhair}
\end{figure}

While the mass of the Schwarzschild black hole can be all positive values, new scalar hairy ones emerge only in the restricted mass region. For $\mu=1/100$ and $\beta=50$, we find that $M_{\rm max}=5.505$ and $M_{\rm min}=4.699$. The mass/horizon dependence is plotted in the left panel of Fig.~\ref{mrscalarhair}. It is of interest to note that for the same $r_+$, the new black hole has bigger mass than that of the Schwarzschild, indicating the condensation of the scalar hair contributes to the mass. As a contrast, the mass would be smaller when the black hole carries the ghostlike massive spin-2 hair \cite{Lu:2015cqa}.

The black hole solutions have three parameters, the couplings $(\mu,\beta)$ and mass $M$, which are all dimensionful.  We can form dimensionless parameters $\tilde M=M/\sqrt{\beta}$, $\tilde \mu = \mu\sqrt{\beta}$ or $\hat \mu =\mu M$.  All dimensionless quantities of the black hole must be functions of either the pair $(\tilde M, \tilde \mu)$ or the alternative but equivalent choice $(\tilde M, \hat \mu)$. The dimensionless scalar hair parameter $\phi_+$ on the horizon and $\phi_0/\sqrt{\beta}$ of the asymptotic are shown in the right panel of Fig.~\ref{mrscalarhair}.

\begin{figure}[htp]\label{entropytemp}
\begin{center}
\includegraphics[width=200pt]{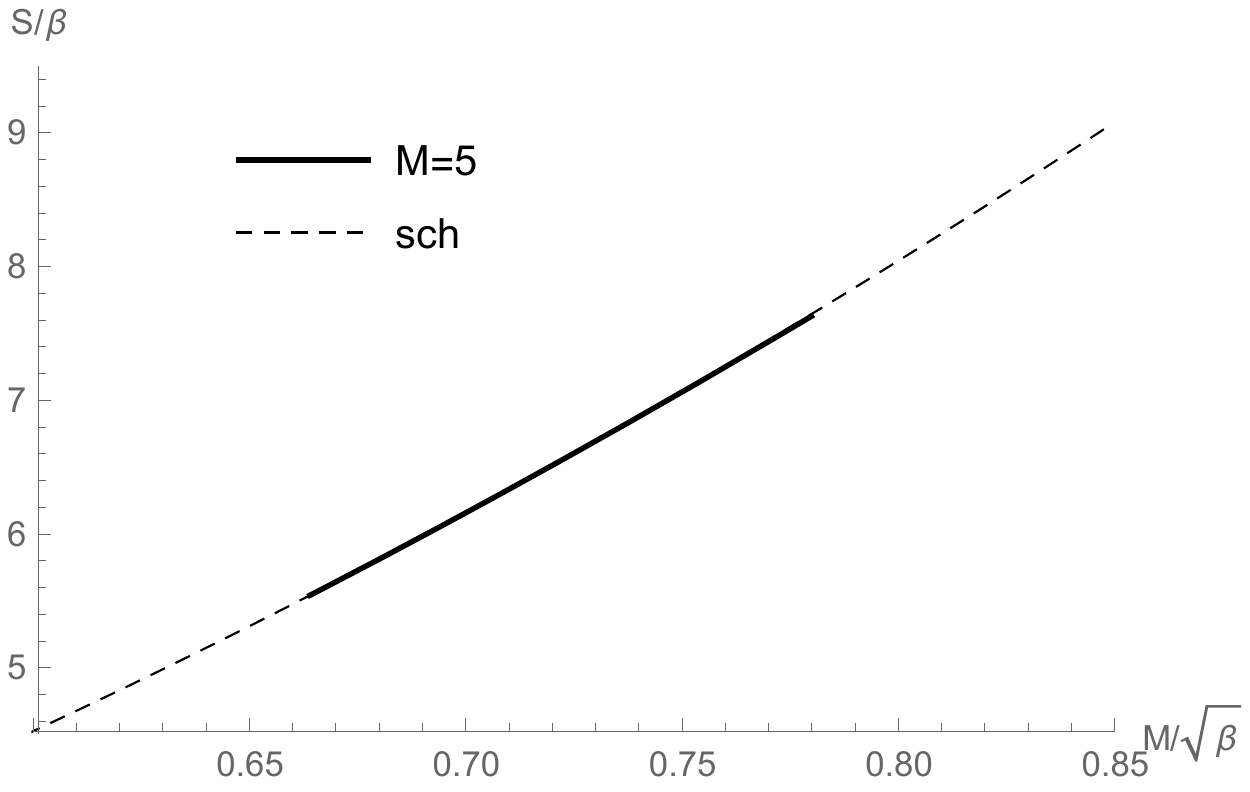}\ \ \ \ \
\includegraphics[width=200pt]{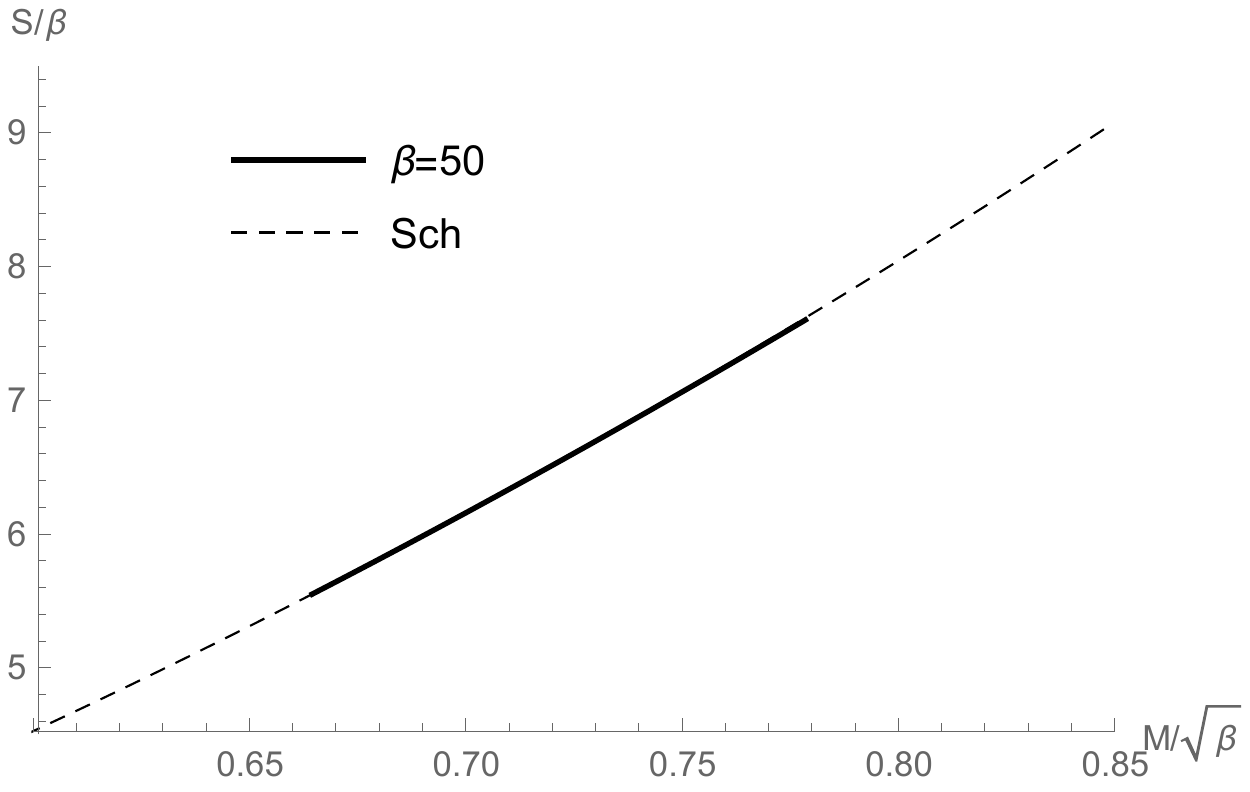}
\end{center}
\caption{\small\it Both show the (dimensionless) entropy/mass relation for the new hairy black holes, indistinguishable from that of the Schwarzschild, for the $\mu=1/100$. The left panel keeps $\hat \mu$ fixed, whilst the right panel keeps $\tilde \mu$ fixed.
}
\end{figure}

We now examine the black hole thermodynamics for the new hairy solutions.  It should be pointed out that the massive scalar hair $\phi_0$ cannot enter the first law of black hole thermodynamics since its thermodynamical conjugate is associated with divergent mode in the solutions and set to zero. Consequently the first law remains $dM=T dS$. The intriguing property of our new solutions is that for given mass $M$, although the horizon radius is smaller than the Schwarzschild black hole, the entropy and hence temperature appear to be the same as those of the Schwarzschild, as shown in Fig.~\ref{entropytemp}.  In other words, although the new hairy black hole is very different geometrically from the Schwarzschild, the thermodynamical properties are the same.

\begin{figure}[htp]
\begin{center}
\includegraphics[width=150pt]{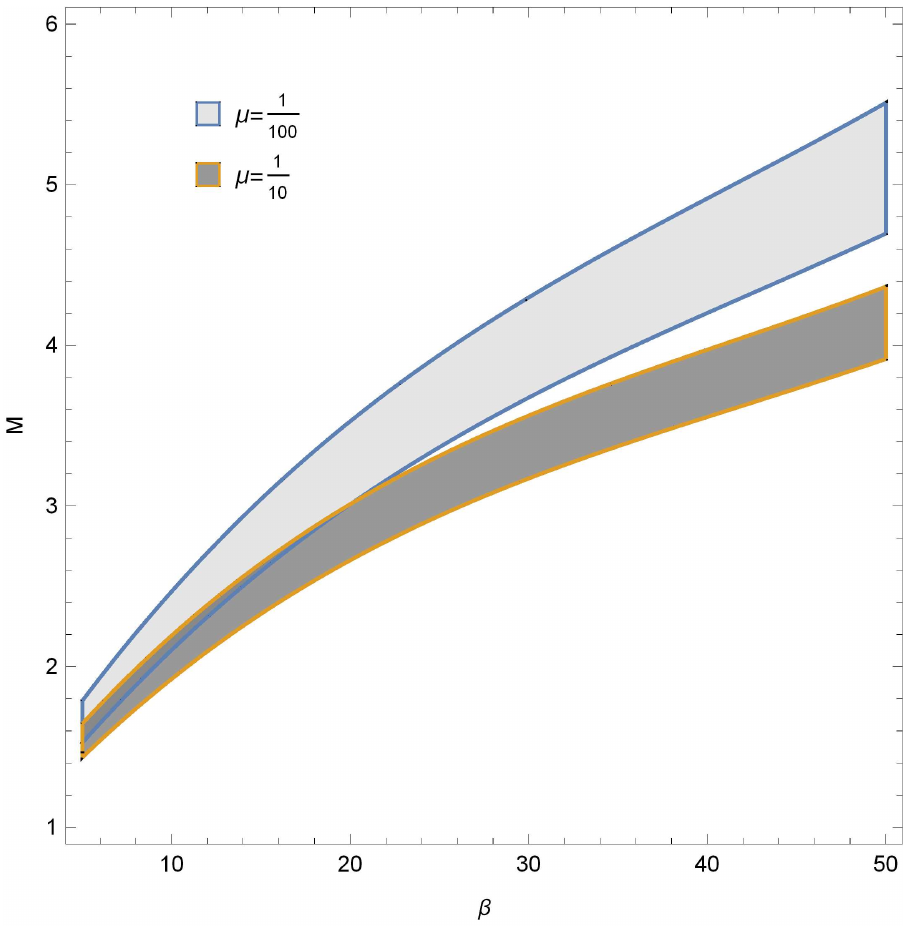}
\end{center}
\caption{\small\it The panel shows the ranges of $M$ and $\beta$ where black hole scalarization occurs for certain given $\mu$'s. }
\label{massrange}
\end{figure}

We have analysed a wide range of parameter space of the couplings $(\mu,\beta)$ and the properties described above hold in general.  The allowed mass ranges for the new hairy black holes depend on the couplings $(\mu,\beta)$.  For $\mu=1/10$ and $1/100$, we plot the mass range dependence on $\beta$ in Fig.~\ref{massrange}. Data fitting up to the cubic order gives (for $\beta\ge5$,)
\bea
\mu=\ft{1}{100}:&& M_{\rm max} \approx 0.985 + 0.174\beta -0.00278 \beta^2 + 0.0000223 \beta^3\,,\nn\\
&&M_{\rm min} \approx 0.848 + 0.146\beta - 0.00226\beta^2 + 0.0000175\beta^3\,,\nn\\
\mu=\ft{1}{10}:&& M_{\rm max} \approx 0.984 + 0.145\beta - 0.00257\beta^2 + 0.0000207\beta^3\,,\nn\\
&&M_{\rm min}\approx 0.850 + 0.127\beta - 0.00219\beta^2 + 0.0000173\beta^3\,.
\eea
It shows for given $\mu$, the bigger $\beta$ gives bigger the mass range and for given $\beta$, the smaller $\mu$ yields the bigger range.  We verify the degeneracy of black hole thermodynamics of the new and the Schwarzschild for all these black holes. Our numerical technique allows us to obtain reliably the solutions with $\beta\ge 5$. For smaller $\beta$, the equations become too unstable to get trustworthy data.

\section{Conclusions}

In this paper, we propose a class of higher-derivative extensions to Einstein gravity, constructed from the Ricci scalar and the Gauss-Bonnet combination.  The theories are ghost free and the linear spectrum contains a massive scalar mode, in addition to the usual graviton. We focus on the simplest such examples, namely the Gauss-Bonnet extensions of the Starobinsky model.  The extensions allow us to overcome the black hole no-hair theorem in the Starobinsky model and construct new hairy black holes for some restricted range of black hole mass. Our result indicates that black hole scalarization should be a common phenomenon in general higher-derivative gravities involving massive scalar modes. Owing to the massiveness, these black hole scalarization are invisible in the long range, but should play important roles in quantum gravity and early cosmology.  Our numerical results also indicate degeneracy of black hole thermodynamics.

It is worth noting that for the Gauss-Bonnet extended Starobinsky model (\ref{extension2}) that we focused on, in addition to the Minkowski vacuum, there are two de Sitter vacua for positive $\beta$ and two anti-de Sitter vacua for negative $\beta$. It is worth investigating the black hole scalarization in these backgrounds as well. Our construction can also be easily generalized to general $f(R)$ gravity, by considering
\be
{\cal L}=\sqrt{-g} \Big(\Phi R - V(\Phi) + U(\Phi) E^{\rm GB}\Big)\,.
\ee
Integrating out the algebraic $\Phi$ gives rise a class of $f(R,E^{\rm GB})$ gravities and the Gauss-Bonnet extended Starobinsky model is the simplest example.  It is of great interest to study the general aspects of black hole scalarization and also implications in cosmology.

\section*{Acknowledgement}

H.S.~Liu is supported in part by NSFC (National Natural Science Foundation of China) Grant No.~11675144.,  H.~L\"u is supported in part by NSFC Grants No. 11875200 and No. 11935009. Z.Y.~Tang and B.~Wang are supported in part by NSFC Grants.

\end{document}